\documentclass{article}
\usepackage[a4paper, total={6in, 10in}]{geometry}
\usepackage{graphicx} 
\usepackage{appendix}
\usepackage{hyperref}
\usepackage{authblk}
\hypersetup{
    colorlinks=true,
    linkcolor=blue,
    filecolor=magenta,      
    urlcolor=blue,
    pdftitle={US Early Career ESG Input},
    pdfpagemode=FullScreen,
    }

\begin{document}

\title{United States Early Career Researchers in Collider Physics input to the European Strategy for Particle Physics Update}
\date{March 2025}
\author[1]{Summary prepared by Oz Amram\thanks{oz.amram@cern.ch}}
\author[1]{Grace Cummings\thanks{gcumming@fnal.gov}}
\affil[1]{Fermi National Accelerator Laboratory}

\maketitle
\begin{abstract}
This document represents a contribution of the United States early career collider physics community to the 2025--2026 update to the European Strategy for Particle Physics. 
Preferences with regard to different future collider options and R\&D priorities were assessed via a survey. 
The early career community was defined as anyone who is a graduate student, postdoctoral researcher, untenured faculty member, or research scientist under 40 years of age.
In total, 105 participants responded to the survey between February and March 10th, 2025.  Questions were formulated primarily to gauge the enthusiasm and preferences for different collider options in line with the recommendations of the United States' P5 report, relevant to the European Strategy Update. 
\end{abstract}

\newpage

\section{Introduction}

A key outcome of the current update to the European Strategy is a recommendation for the next energy frontier collider project hosted at CERN. 
Given the grand scale and long timeline of any proposed future collider, the decision will shape the entire field of particle physics for the next several decades.
The careers of particle physicists in the United States will undoubtedly be greatly affected by the outcome of this process.
Early career collider physicists in particular have a lot at stake, as the majority of our career will occur after the conclusion of High Luminosity LHC, and thus will be determined by whatever project is built next. 
Historically, the United States has had significant involvement in the physics program at CERN, and given this history, it is fair to anticipate this opportunity again.

The US particle physics community has recently come together in the Snowmass--P5 process to decide on a vision and prioritization for the next decade of research, culminating in the widely accepted P5 report\footnote{\url{https://www.usparticlephysics.org/2023-p5-report/}}. 
In relation to future collider physics programs, a key area of the current European Strategy Update, the P5 report made several statements. 
The report recommended US participation in an off-shore Higgs Factory, and research and development toward a parton-center-of-mass collider of 10 TeV.
Within this guidance, the P5 report could support several different collider options at CERN. 
While the Snowmass--P5 process received input from early career scientists, the final report synthesizes views from the entire community, and does not specifically highlight the opinions of the early career community.  
Given the long term impact of the current update to the European Strategy, and the generality of the P5 report, we felt it was a useful exercise to directly assess the enthusiasm of the US early career community for different collider options consistent with the recommendations of the P5 report. 

A survey was designed to assess different proposed colliders. 
Respondents were also asked to consider R\&D priorities and communicate preferences for different long-term options beyond the next immediate collider. 
The survey did not explicitly ask about the perception of technical and financial feasibility of the different projects, but these factors likely played a role in the project preferences.
The survey was distributed to various experimental collaborations (ATLAS, CMS, Belle-II, LHCb, ALICE) through dedicated early career mailing lists when possible, the US LHC Users Association, and among theorists. Responses were solicited February 10th to March 10th, 2025. 
Any early career researcher associated with the United States was encouraged to submit input. For the purposes of this survey, early career was defined as anyone who is a graduate student, postdoctoral researcher, untenured faculty member, or research scientist under 40 years of age. 
A total of 105 responses were collected.
The survey was anonymous, but participants could optionally enter their emails if they wished to provide feedback on the summary document. 

\section{Survey Results}
In this section we summarize the survey responses. 
The exact distributions for each question are given in Appendix \ref{sec:figs_add}.
The anonymized survey responses have been made public for transparency \footnote{\url{https://docs.google.com/spreadsheets/d/1oeMfGNbFBOYMDkJyZsCHbP48nKHzt1BbtG9XqZelfRY}}.

Given the statistical power of the 105 respondents, the uncertainty on the mean value of the questions asking for a 1 through 5 rating is ${\sim}0.1$.
The uncertainty on the outcome yes/no poll questions asked is ${\sim}5\%$.
Many questions had significant heterogeneity in the responses.
Given the limited range of the rating scale we do not report the standard deviation of the rating scores but instead assess the homogeneity of responses based on the fraction of responses in the 1-2 range vs. 4-5 range. 

\subsection{Respondent Demographics}
\begin{figure}
    \centering
    \includegraphics[width=0.45\linewidth]{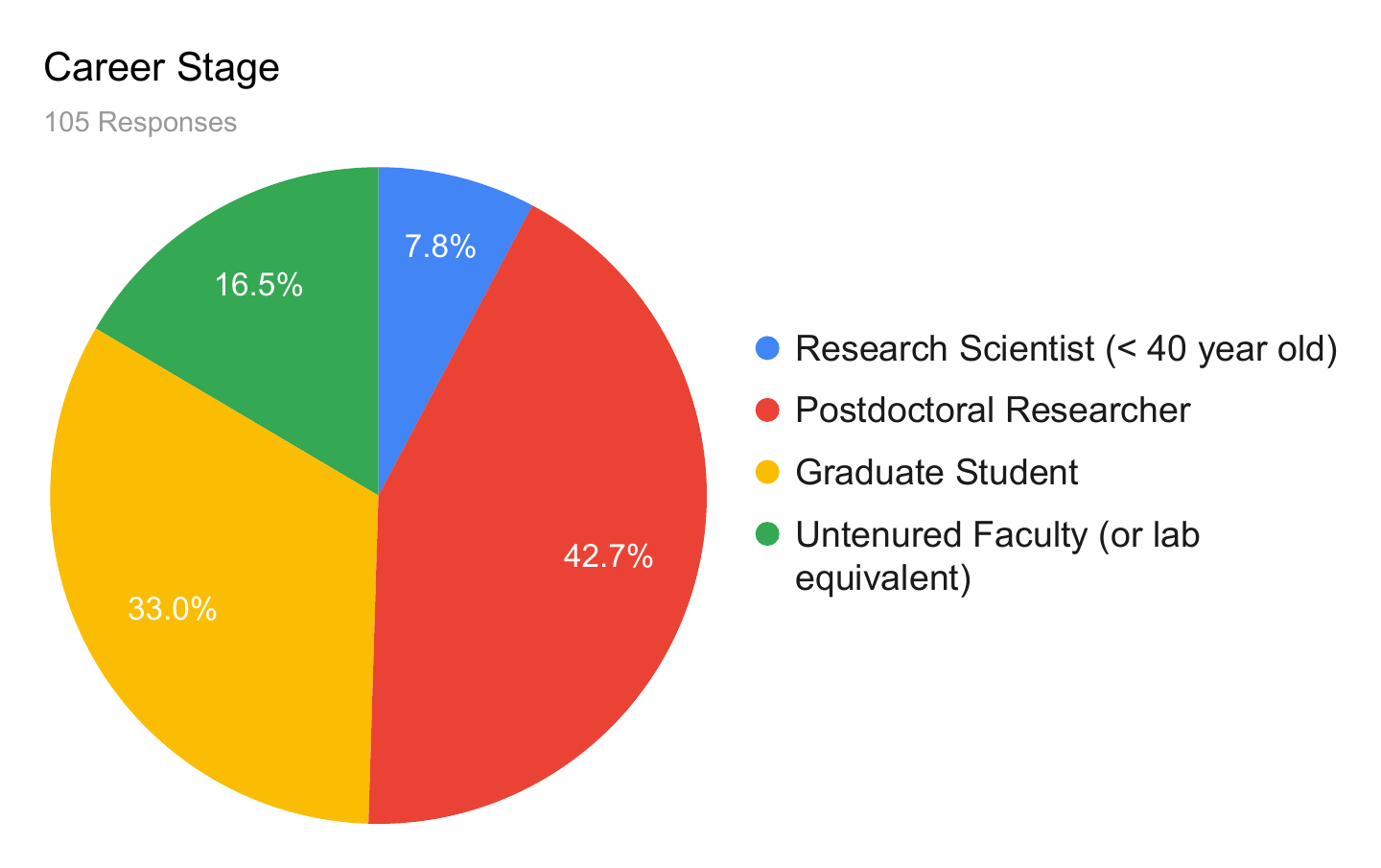}
    \includegraphics[width=0.45\linewidth]{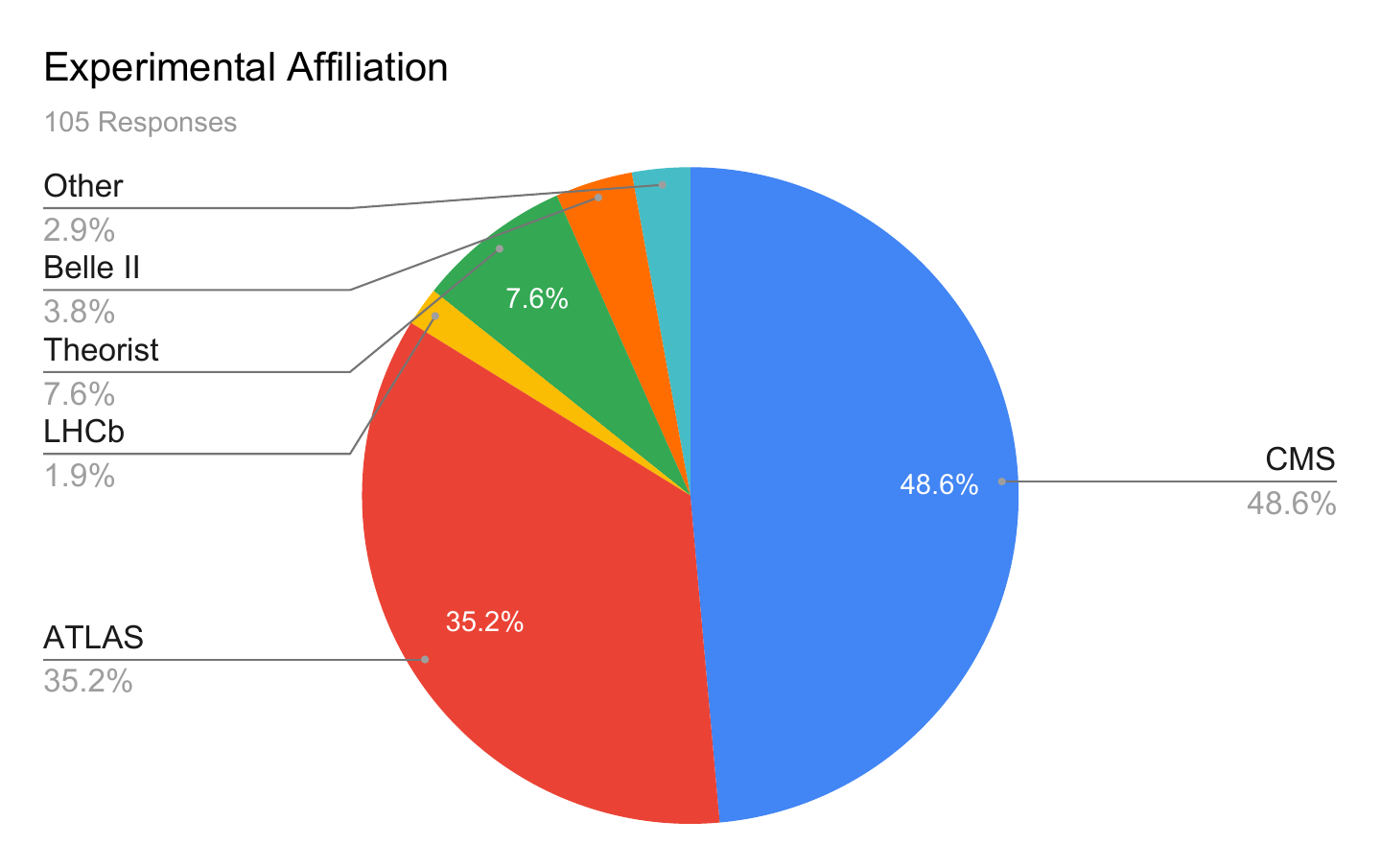}
    \caption{Left: Summary of the self-reported career stages of the respondents of this survey. Right: Experimental affiliation or field of study of respondents of this survey.}
    \label{fig:demo_sum}
\end{figure}

Of the 105 survey respondents, the largest career stage represented was postdocs (43\%), followed by graduate students (32\%), untenured faculty (17\%) and research scientists (8\%). 
The majority of survey respondents were members of a current LHC collaboration, with 47\% and 35\% of respondents part of the CMS and ATLAS Experiments, respectively. Theorists made up 
8\% of respondents, and the remaining 10\% came from various other particle physics experiments (ALICE, LHCb, Belle-II, and others). Figure \ref{fig:demo_sum} summarizes the career stage and experimental affiliation of the respondents of this survey. Primary work location was also surveyed, with 54\% of the respondents based at a US University, 25\% at a US national laboratory, 20\% at CERN, and the remaining 1\% at other European institutes. Additional demographic figures can be found in Appendix \ref{sec:figs_adddemo}.

\subsection{Project Enthusiasm}
The first section of the survey assessed excitement for the physics opportunities at various collider options, asking participants to rate their enthusiasm for each project on a scale of 1 (not excited at all) to 5 (extremely excited). 
The collider options offered for consideration were the FCC-ee, a linear electron-positron Higgs factory (such as CLIC or the ILC), an immediate FCC-hh at energy of ${\sim}85$ TeV, beginning operation in ${\sim}2050$, a 3 TeV muon collider, and a 10 TeV muon collider. Respondents were encouraged to write-in collider scenarios not covered. 
The enthusiasm of the different physics cases can be summarized as follows:

\begin{itemize}

\item The FCC-ee was given an average score of 3.2. There was large heterogeneity between responses, with 30\% giving a 1 or 2, and 42\% giving a 4 or 5. 

\item A linear electron-positron Higgs factory received a similar average rating of 3.2, with similar heterogeneity.

\item The proposal for an 85 TeV FCC-hh starting in $\sim$~2050 had a slightly higher average rating of 3.7. 
There was moderate heterogeneity in responses, with 18\% giving a 1 or 2 and 66\% giving a 4 or 5. 

\item The two muon collider options presented, 3 TeV and 10 TeV, received higher average scores of 4.1 and 4.6 respectively. There was minimal heterogeneity in responses, with 4\% giving a 1 or 2 and 87\% giving a 4 or 5. 
\end{itemize}

\begin{figure}
    \centering
    \includegraphics[width=0.7\linewidth]{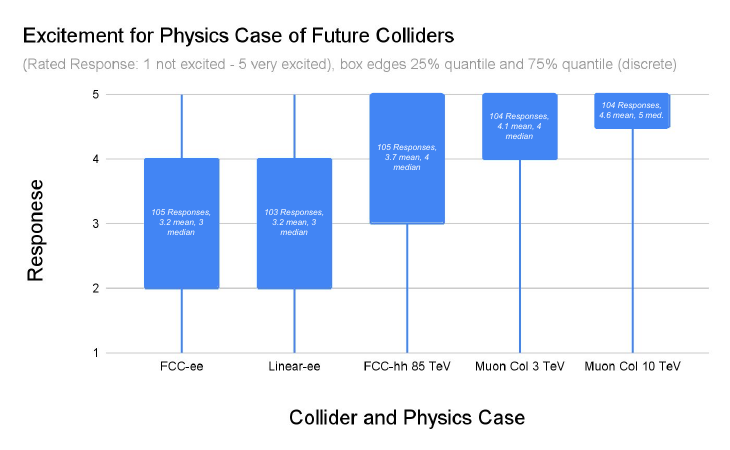}
    \caption{Summary of excitement-level for the physics cases of different prospective colliders. Participants rated their enthusiasm for each project on a scale of 1 (not excited at all) to 5 (extremely excited). The filled boxes represent the range of the distribution between the 25th and 75th percentiles, and the whiskers represent the minimum and maximum. The percentile values reflect the discrete nature of the data.}
    \label{fig:excite_sum}
\end{figure}

Common write-in options were the HE-LHC and LEP3, as well as as plasma-wakefield colliders such as HALHF were mentioned. Collider programs with similar physics program, like the CEPC, were also mentioned. Figure \ref{fig:excite_sum} provides a visual summary of the relative enthusiam-levels for the different physics cases. Appendix \ref{sec:figs_addexcite} contains the full distributions for each collider scenario.

\subsection{Project Staging}

The next thrust of the survey assessed enthusiasm for different project staging scenarios. The same 1 to 5 scale was used,  with 1 meaning not excited at all, and 5 being extremely excited. 

\begin{itemize}
\item The integrated FCC-ee program, which would have the FCC-ee starting in $\sim$~2048 and and FCC-hh in $\sim$~2070 was given an average score of 3.1. 
There was large heterogeneity between responses, with 36\% giving a 1 or 2 and 43\% giving a 4 or 5. 

\item The program of a linear Higgs factory followed by FCC-hh was given an average score of 3.2. 
There was large heterogeneity between responses, with 32  \% giving a 1 or 2 and 44\% giving a 4 or 5. 

\item The program of FCC-ee followed by a muon collider was given an average score of 3.3. 
There was moderate heterogeneity between responses, with 25\% giving a 1 or 2 and 45\% giving a 4 or 5. 

\item The program of linear Higgs factory followed by a muon collider was given an average score of 3.5. 
There was moderate heterogeneity between responses, with 24\% giving a 1 or 2 and 57\% giving a 4 or 5. 

\end{itemize}

Participants were also asked to rate how much their enthusiasm for each project would increase if timelines were accelerated such that the project occurred earlier.
A 1 to 5 scale was used, where 1 indicated that their enthusiasm would not increase at all while 5 indicated their enthusiasm would greatly increase. 

\begin{itemize}

\item FCC-ee and a linear Higgs factor received similar average scores of 3.2. There was large heterogeneity for both, with responses being nearly equally split between the five different ratings. 

\item FCC-hh had a higher average score of 3.9. There was mild heterogeneity, with 85\% of respondents rating a 3 or higher. 

\item The muon collider had the highest average score of 4.5. There was minor heterogeneity, 89\% of respondents rated it a 4 or 5. 

\end{itemize}

The individual distributions and project acceleration comparisons are included in Appendix \ref{sec:figs_staging}.

\subsection{Preferences and Research Prioritization}

The next section asked a series of questions about assessing the relative preferences and prioritization of different projects and research goals. Detailed figures are included in Appendix \ref{sec:figs_addpref}. The first series of questions were yes/no answers. 

\begin{itemize}
    \item To the question "\textbf{If China builds the CEPC,} would you prefer a straight to FCC-hh (at 85 TeV) over a Higgs factory option as the next collider at CERN?", a moderate majority of respondents (64\%) answered yes. 
    
    \item To the question "\textbf{If China does not build the CEPC,} would you prefer a straight to FCC-hh (at 85 TeV) over a Higgs factory option as the next collider at CERN?",
    a narrow majority of respondents (57\%) said no. 

    \item To the question "If the US decides to build a muon collider, would you still have interest in working at a CERN-based Higgs factory?" a narrow majority of respondents (57\%) answered yes.

    \item To the question "Would you prioritize R\&D for a multi-TeV muon collider over Higgs factory construction?" a majority of respondents (70\%) answered yes. 

    \item To the question "Would you prioritize accelerated FCC-hh magnet R\&D over Higgs factory construction?" a very narrow majority (52\%) answered yes. 

\end{itemize}

Participants were also asked "How highly would you prioritize plasma wakefield acceleration R\&D (1 being no-interest, 5 being absolutely necessary)".
The average rating given was 3.3. There was moderate heterogeneity between the responses, with 20\% answering 1 or 2 and 46\% answering 4 or 5. 

\subsection{Ranking}
The last section asked participants to rank their preference for 5 different proposed project scenarios.
The five options were "FCC-ee followed by an FCC-hh", "Straight to $\sim$~85 TeV FCC-hh", 
"Linear e+e- machine, with significant muon collider R\&D", "Linear e+e- machine, with significant FCC-hh magnet R\&D", "FCC-ee, with significant muon collider R\&D". 
 
\begin{figure}
    \centering
    \includegraphics[width=0.7\linewidth]{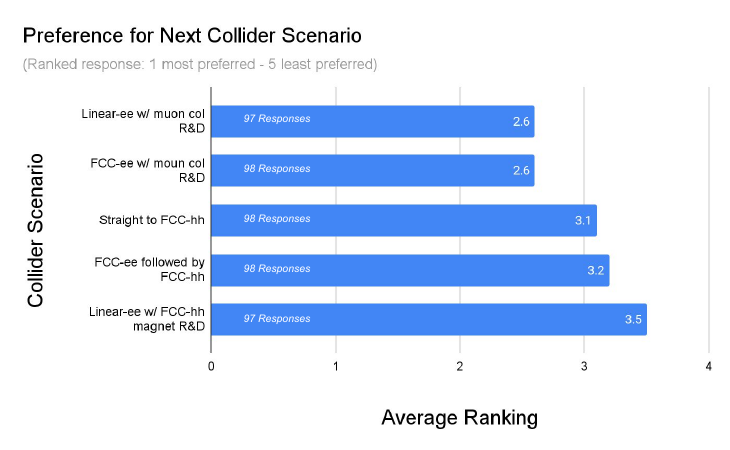}
    \caption{Summary of average rank of each offered collider scenario. Participants ranked collider scenarios from 1 (most preferred) to 5 (least preferred). A lower average ranking thus denotes a more preferred option.}
    \label{fig:rank_sum}
\end{figure}

There was significant heterogeneity in the responses, with all options receiving at least 5 votes as most preferred and at least 8 votes as the least preferred. When computing the average ranking, the most preferred option was given a value of 1, and the least 5, so lower average scores denote more preferred options. Figure \ref{fig:rank_sum} summarizes the results. 

\begin{itemize}
    \item The two options with significant muon collider R\&D, with either a linear Higgs factor or FCC-ee, received the highest preference, with an average ranking of 2.6.
    \item The straight to FCC-hh option received the next highest preference, averaging a rank of 3.1.
    \item The FCC-ee followed by FCC-hh program received an average ranking of 3.2.
    \item A linear Higgs factory with significant FCC-hh magnet R\&D received the lowest average ranking of 3.5. 
\end{itemize}

Running an 'instant runoff' ranked choice voting algorithm\footnote{\url{https://en.wikipedia.org/wiki/Instant-runoff_voting}} on responses led to the FCC-ee with significant muon collider R\&D option narrowly beating out the linear collider with significant muon R\&D option, 53\% to 47\% in the final round). Detailed results for each collider scenario, and additional detailed of the runoff are included in Appendix \ref{sec:figs_addrank}.

\subsection{Free Form Response}
Finally, respondents were invited to contribute written commentary.
Out of the 105 survey participants, 24 included written feedback. 
The comments covered a range of topics, but a common theme was the concern over timelines of different projects.
Several expressed skepticism of the integrated FCC program given that the discovery machine, the FCC-hh, may not occur during their careers. They pointed out it may be difficult to retain the community for this long. 
Several expressed enthusiasm for a muon collider as an exciting project they would like to devote their energy to, while others worried about its technological uncertainties and therefore realistic timeline.

\section{Conclusion}
This report summarizes a survey performed to assess the preferences of the early career collider physics community associated with the United States for the next major collider project at CERN. In total, 
105 responses were received.
The survey did not explicitly ask about the perception of technical and financial feasibility of the different projects, but these factors likely played a role in the project preferences.
A  non-comprehensive list of collider options was assessed by the survey: the FCC-ee, a linear electron-positron Higgs factory, an 85 TeV FCC-hh, and muon colliders.
Based on the survey results, there is significant heterogeneity and diversity of opinion of what is the preferred next project. 
Muon colliders garnered the highest enthusiasm from the surveyed community, and staging scenarios which prioritize R\&D funding for a muon collider in conjunction with a Higgs factory were rated as the most preferred options.
Out of the survey respondents, 70\% preferred to prioritize R\&D for a muon collider over construction of a Higgs factory.
Linear Higgs factory options were given roughly equal ranking to the FCC-ee. 
Between the FCC projects respondents were roughly evenly split between preference for a direct to FCC-hh scenario versus the FCC-ee followed by FCC-hh long term program, even though the physics program of the FCC-hh was rated as more exciting. 
Excitement for all projects increases if they are brought about sooner. 
We hope these preferences will be useful in the European Strategy process. 

\appendix
\section{Figures}
\label{sec:figs_add}
\subsection{Additional Demographic Figures}
\label{sec:figs_adddemo}
\begin{figure}[h]
    \centering
    \includegraphics[width=0.45\linewidth]{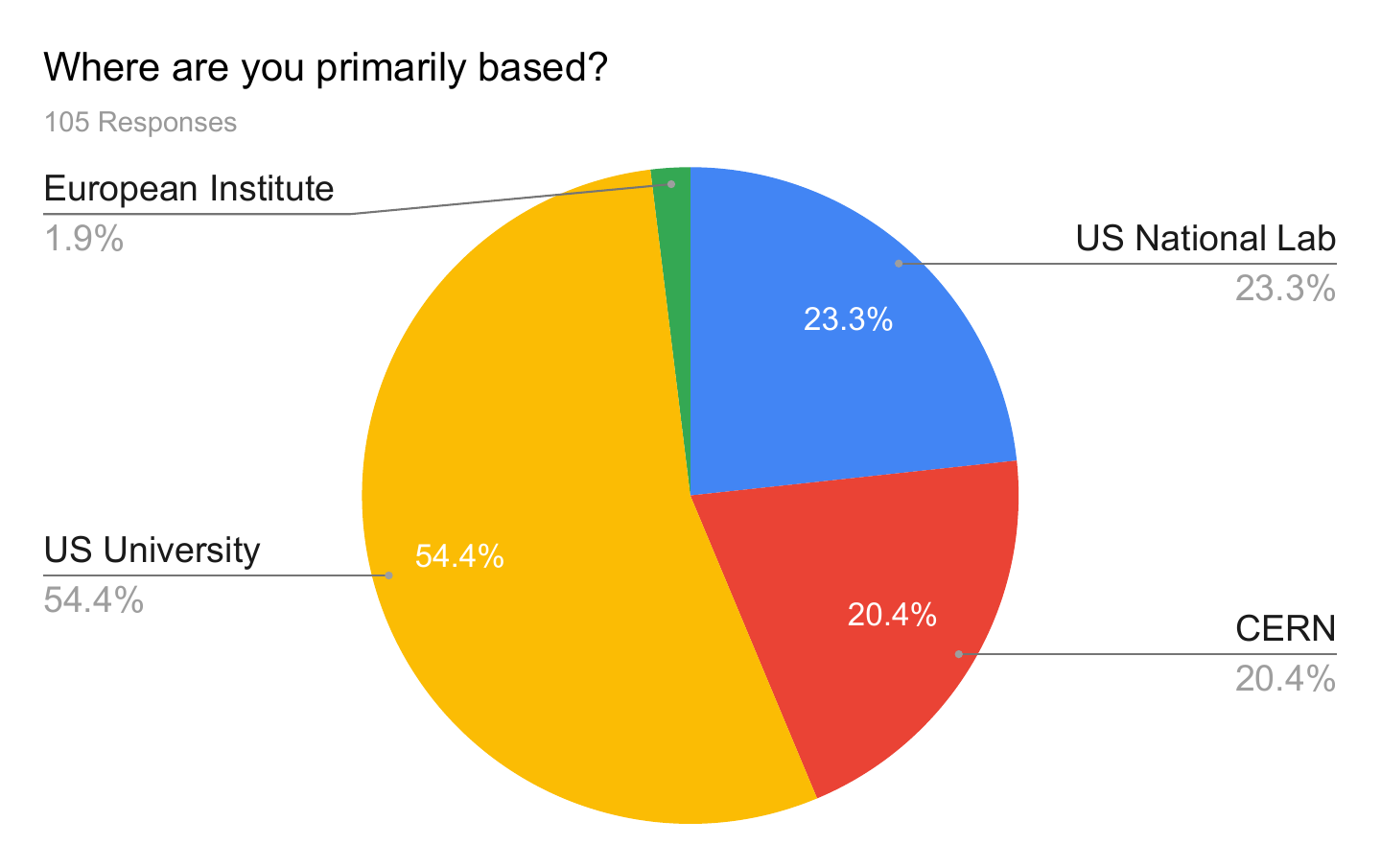}
    \caption{Self-reported primary location of research performed.}
    \label{fig:demos}
\end{figure}
\subsection{Detailed Excitement Figures}
\label{sec:figs_addexcite}
\begin{figure}[h]
    \centering
    \includegraphics[width=0.48\linewidth]{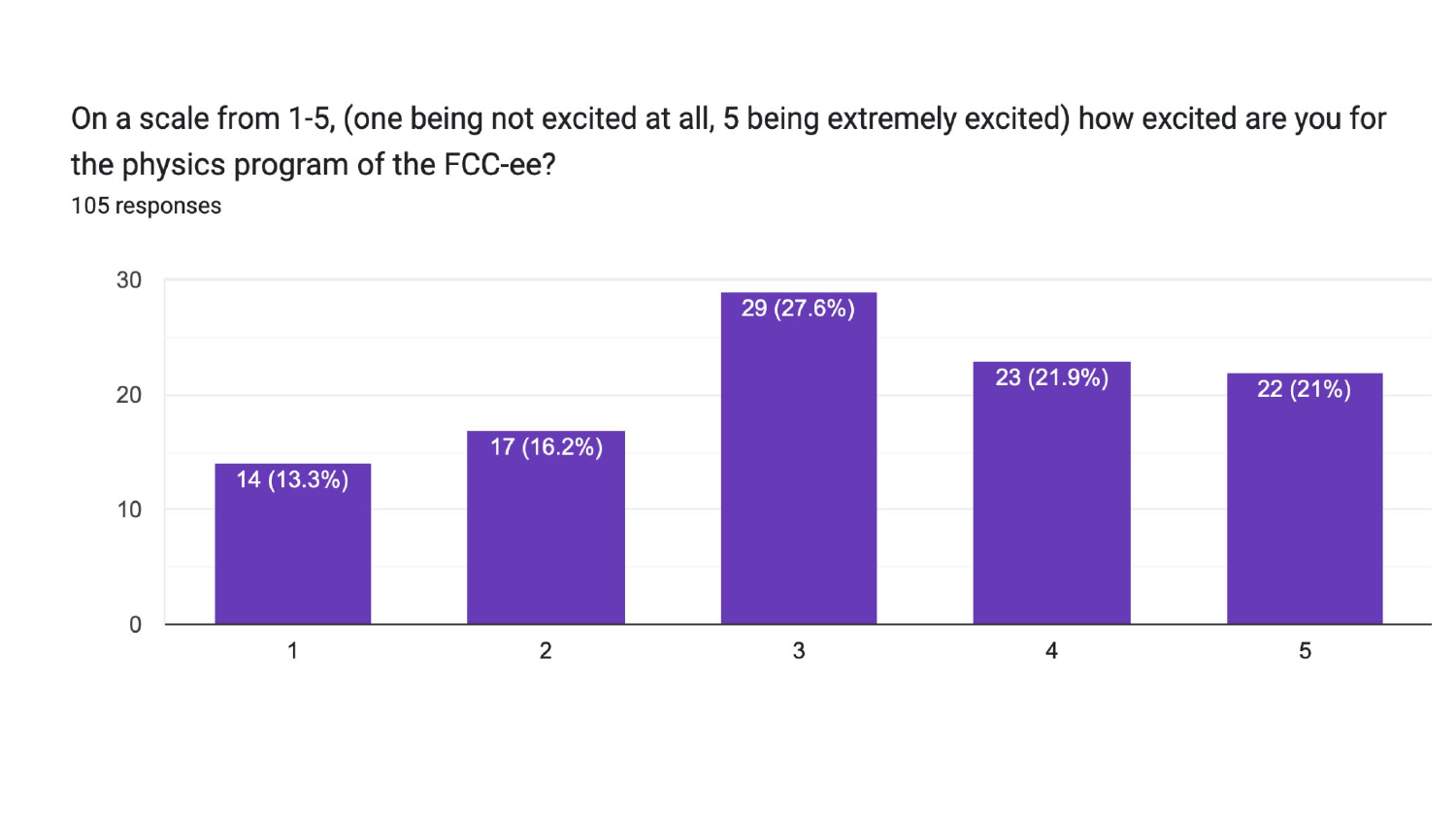}
    \includegraphics[width=0.48\linewidth]{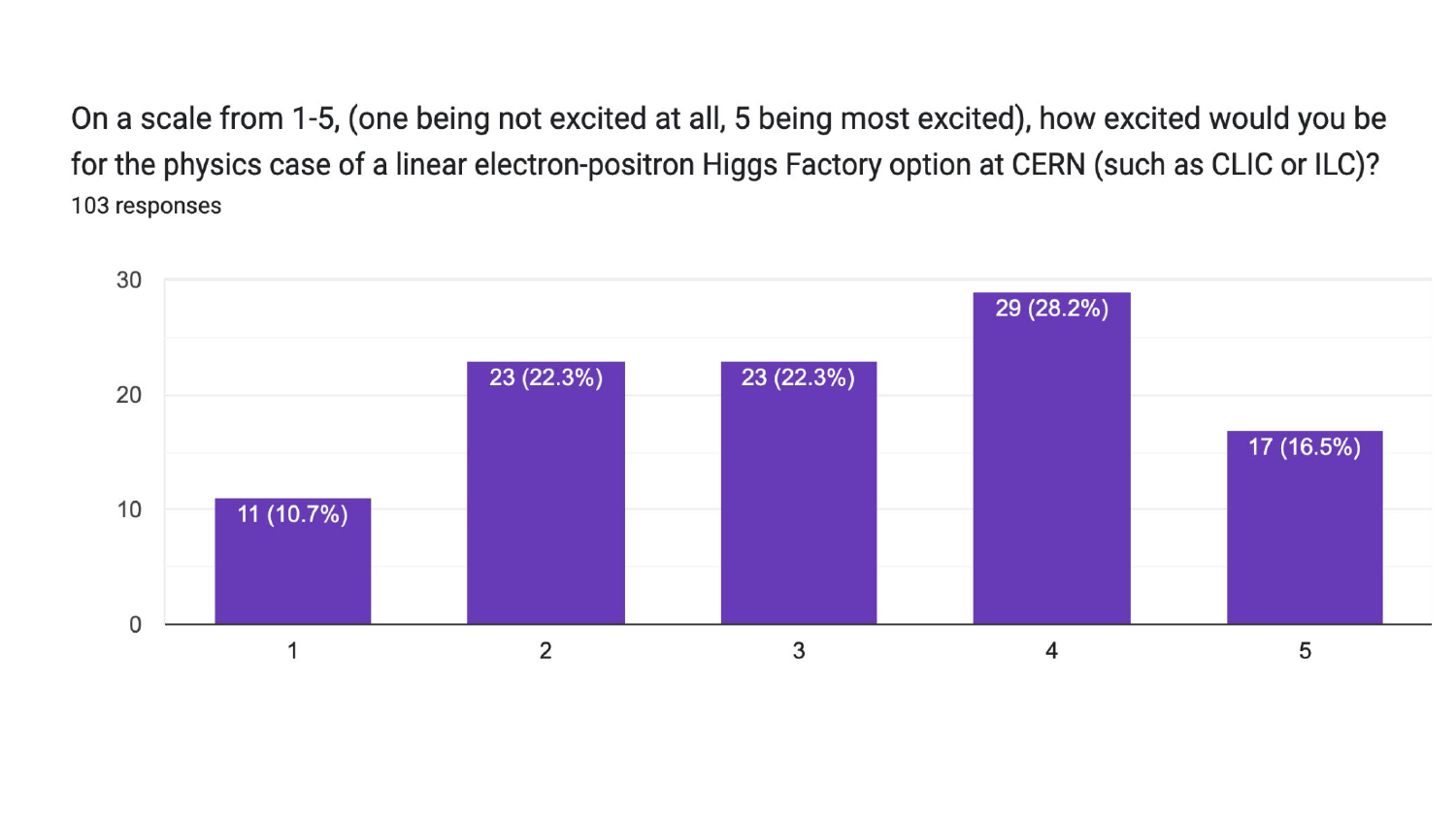}
    \caption{Excitement for different electron/positron collider options. Left: FCC-ee. Right: Linear-ee.}
    \label{fig:excite_collidersee}
\end{figure}

\begin{figure}[h]
    \centering
    \includegraphics[width=0.48\linewidth]{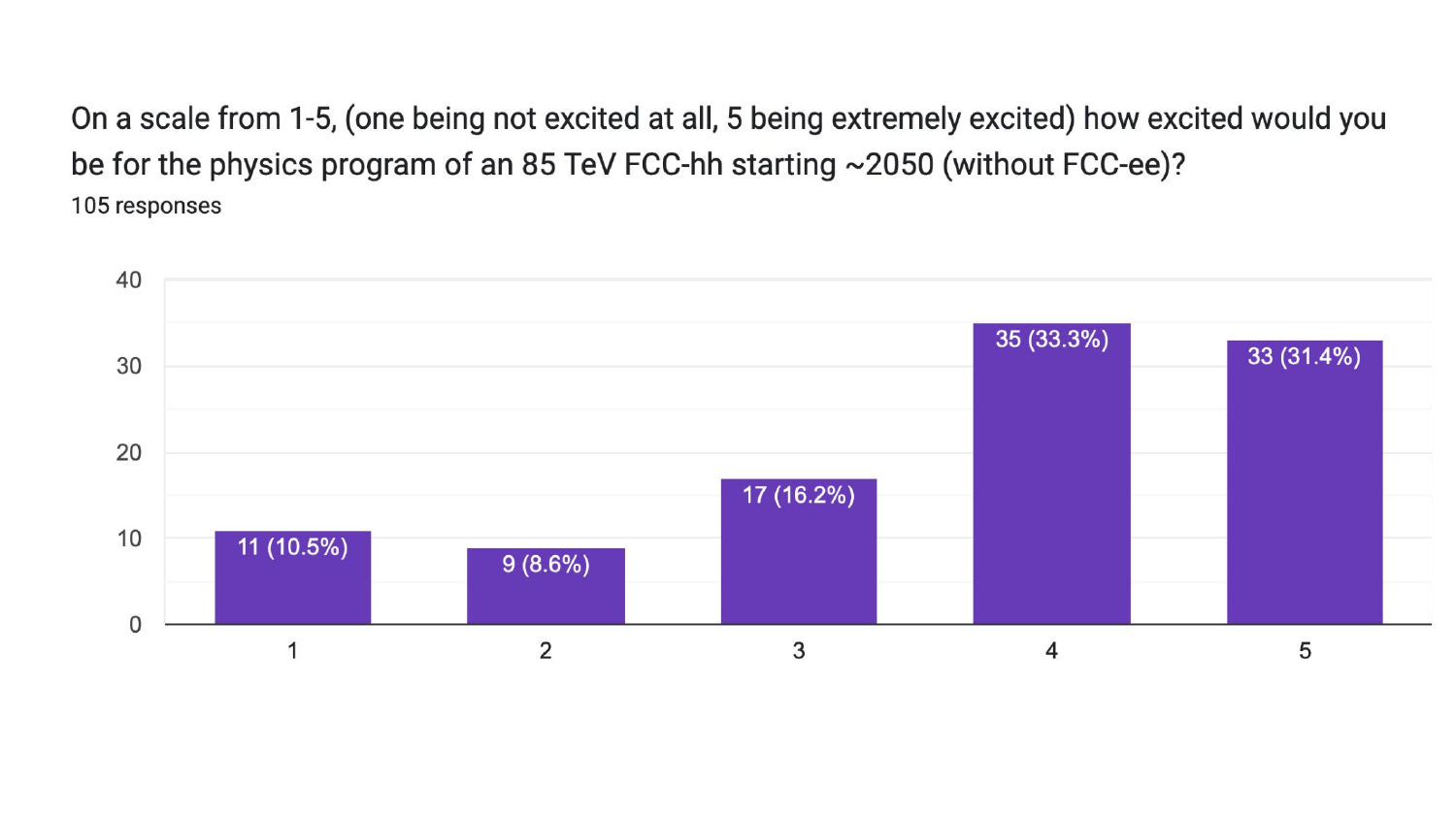}
    \includegraphics[width=0.48\linewidth]{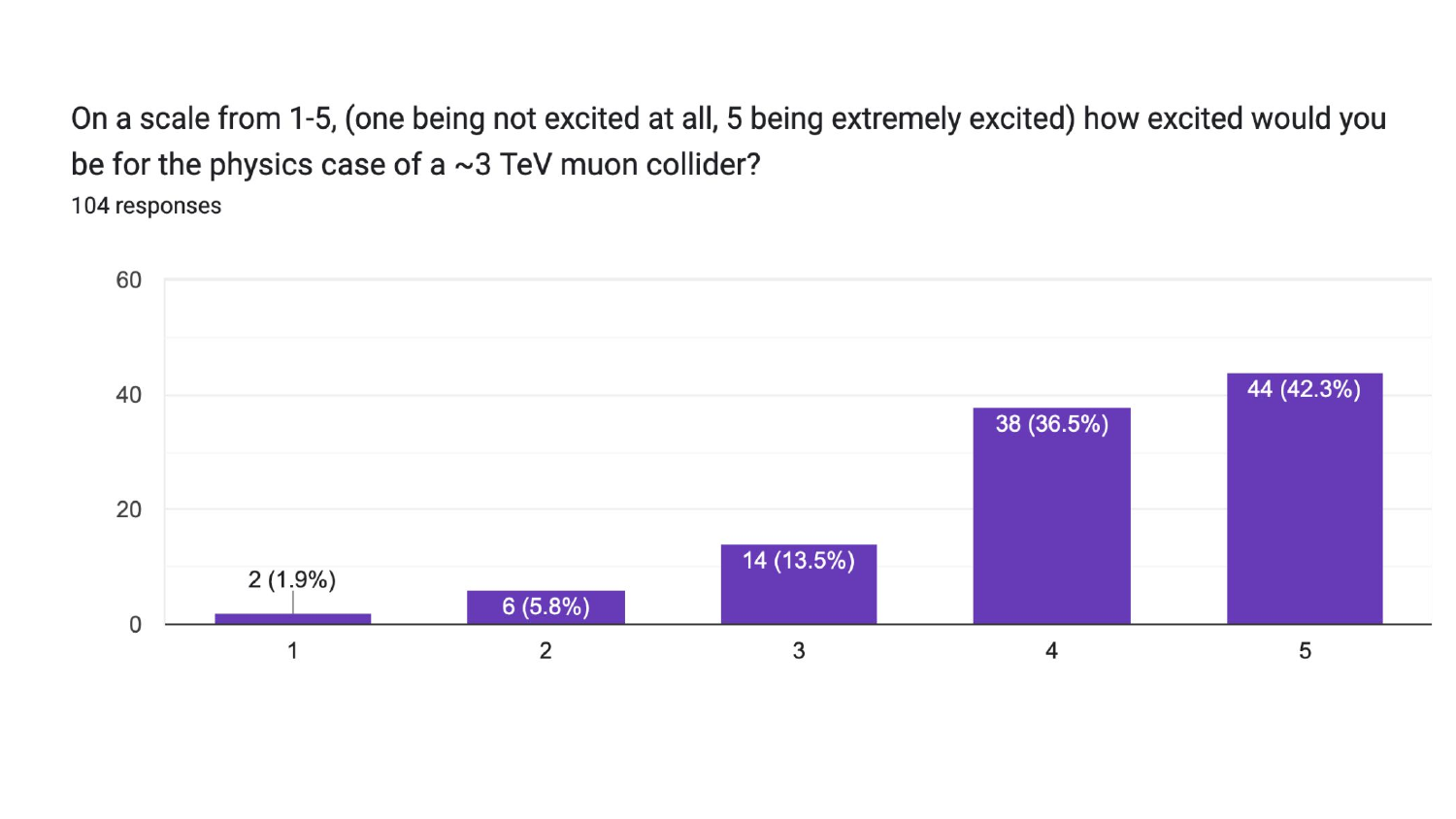}
    \includegraphics[width=0.48\linewidth]{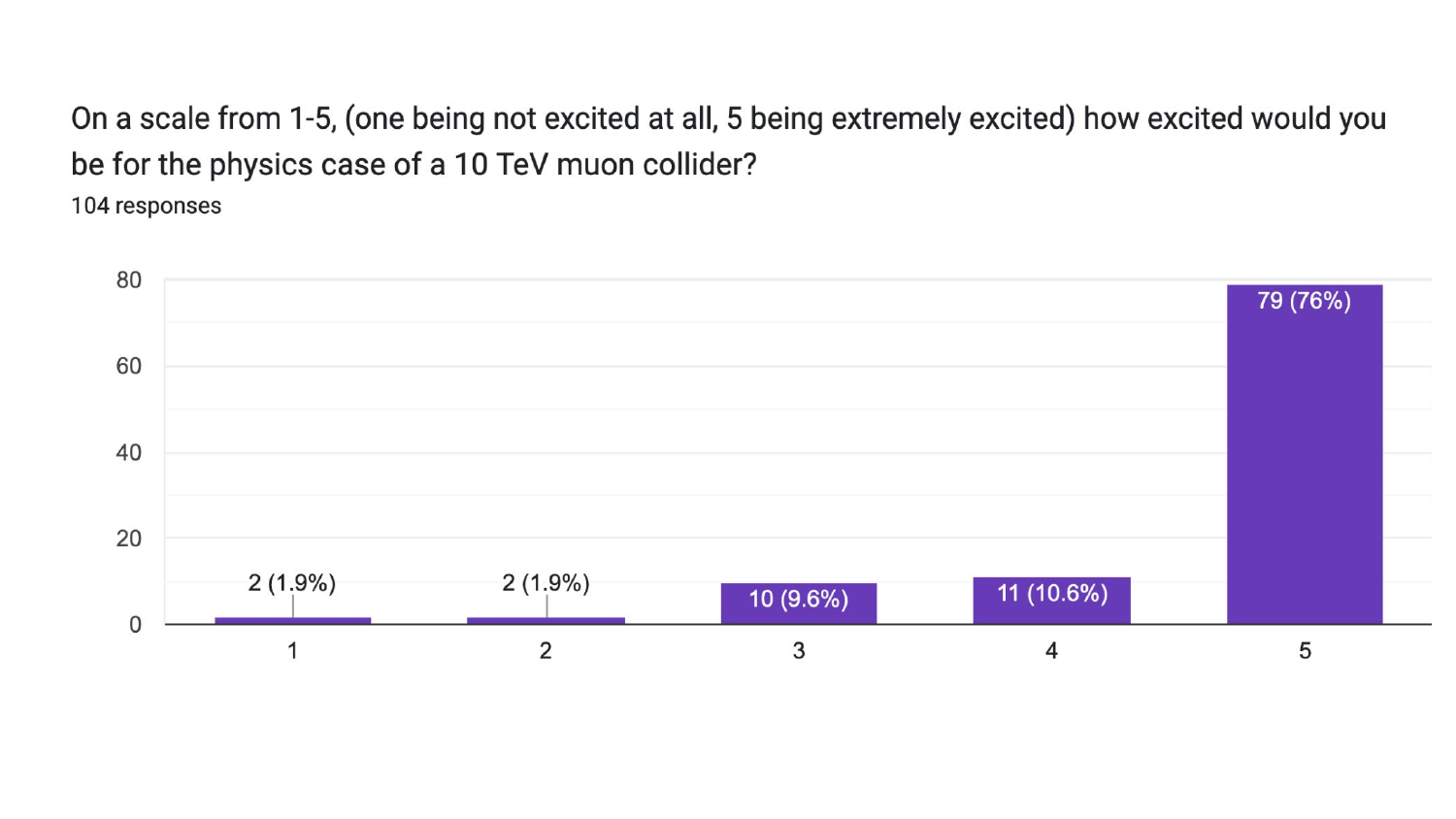}
    \caption{Excitement for higher-energy collider options. Top-left: FCC-hh. Top-right: 3 TeV Muon Collider. Bottom: 10 TeV Muon Collider.}
    \label{fig:excite_colliders}
\end{figure}
\clearpage
\subsection{Project Staging Figures}
\label{sec:figs_staging}

\begin{figure}[h]
    \centering
    \includegraphics[width=0.5\linewidth]{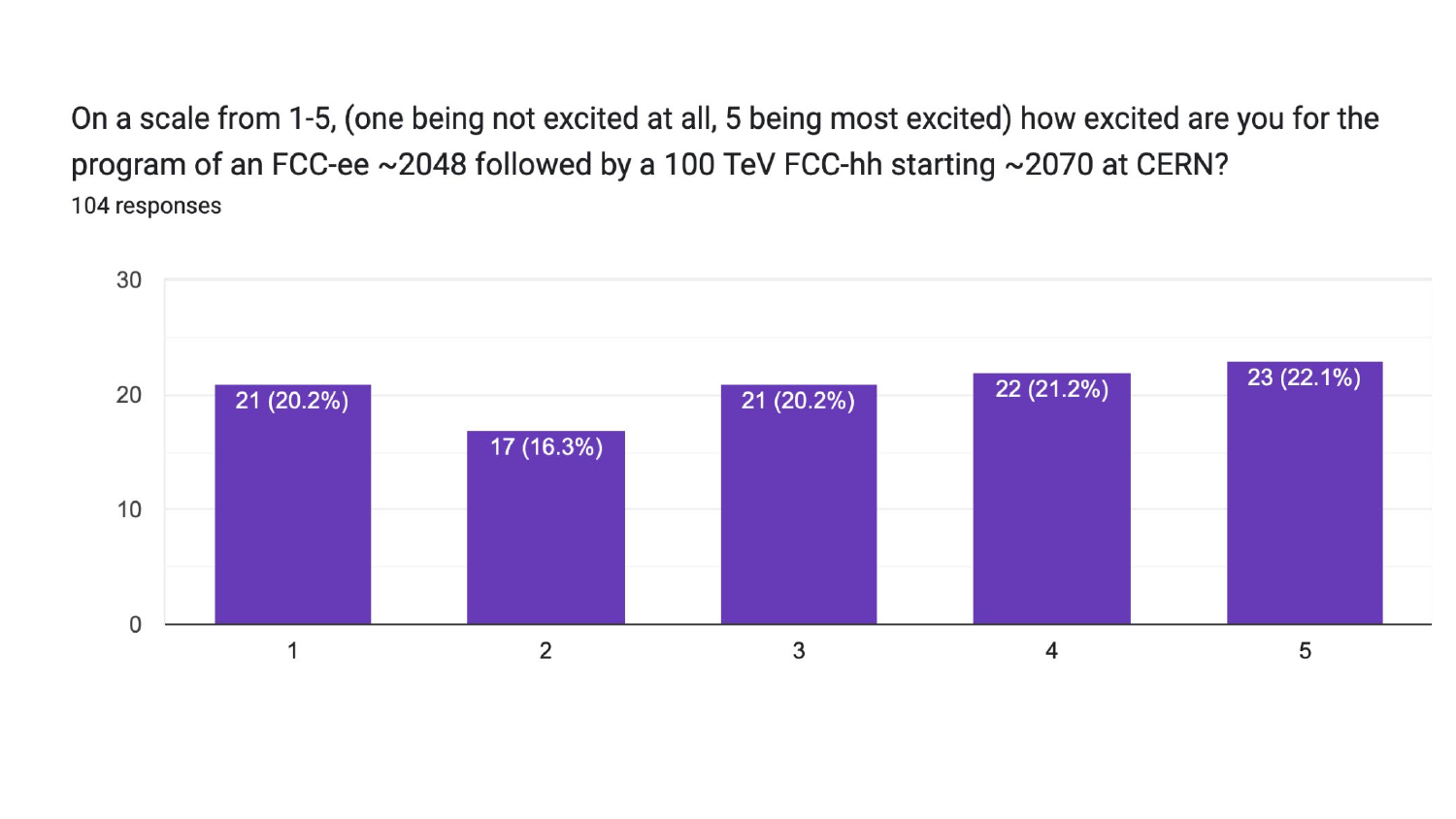}
    \includegraphics[width=0.5\linewidth]{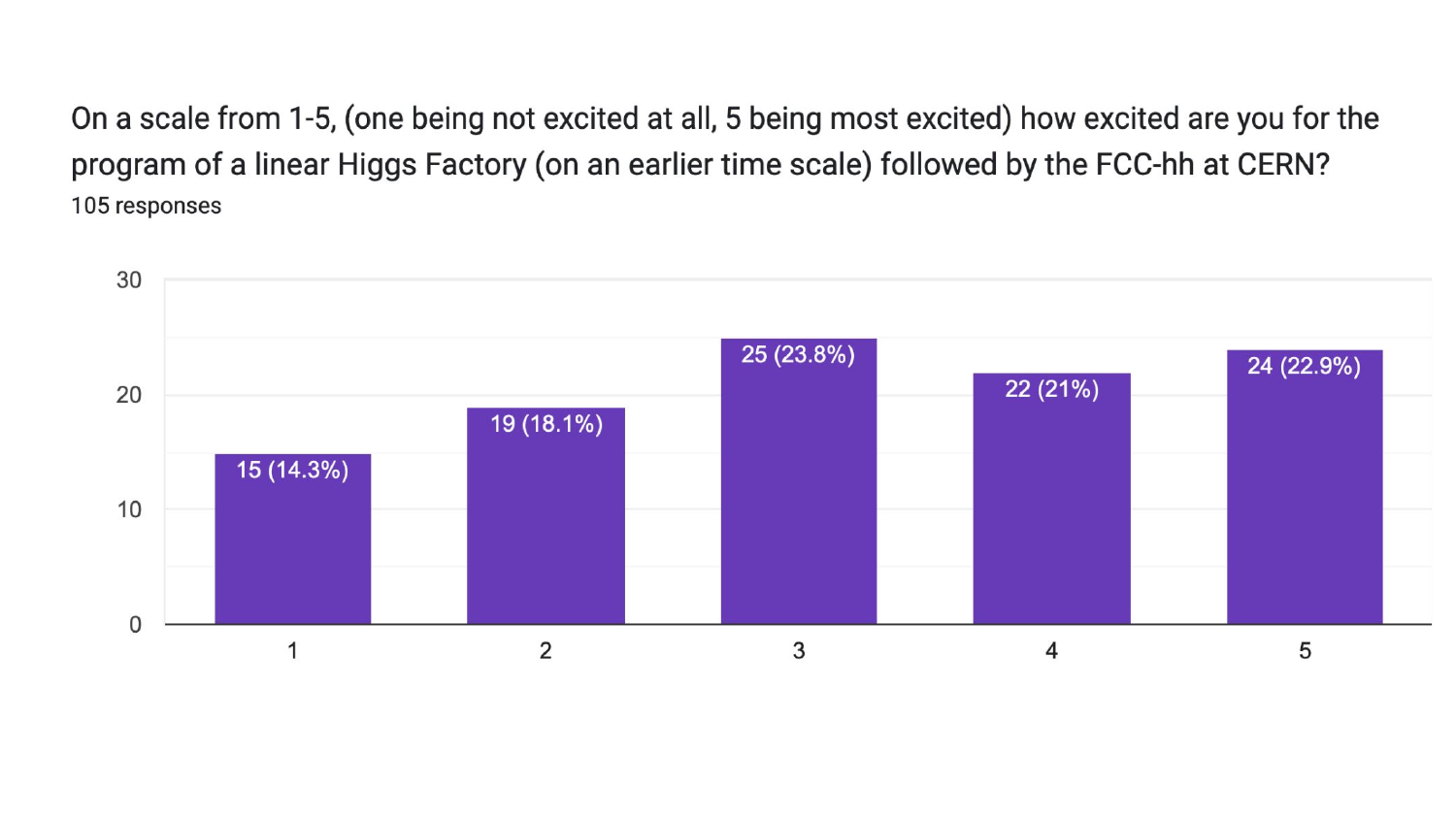}
    \includegraphics[width=0.5\linewidth]{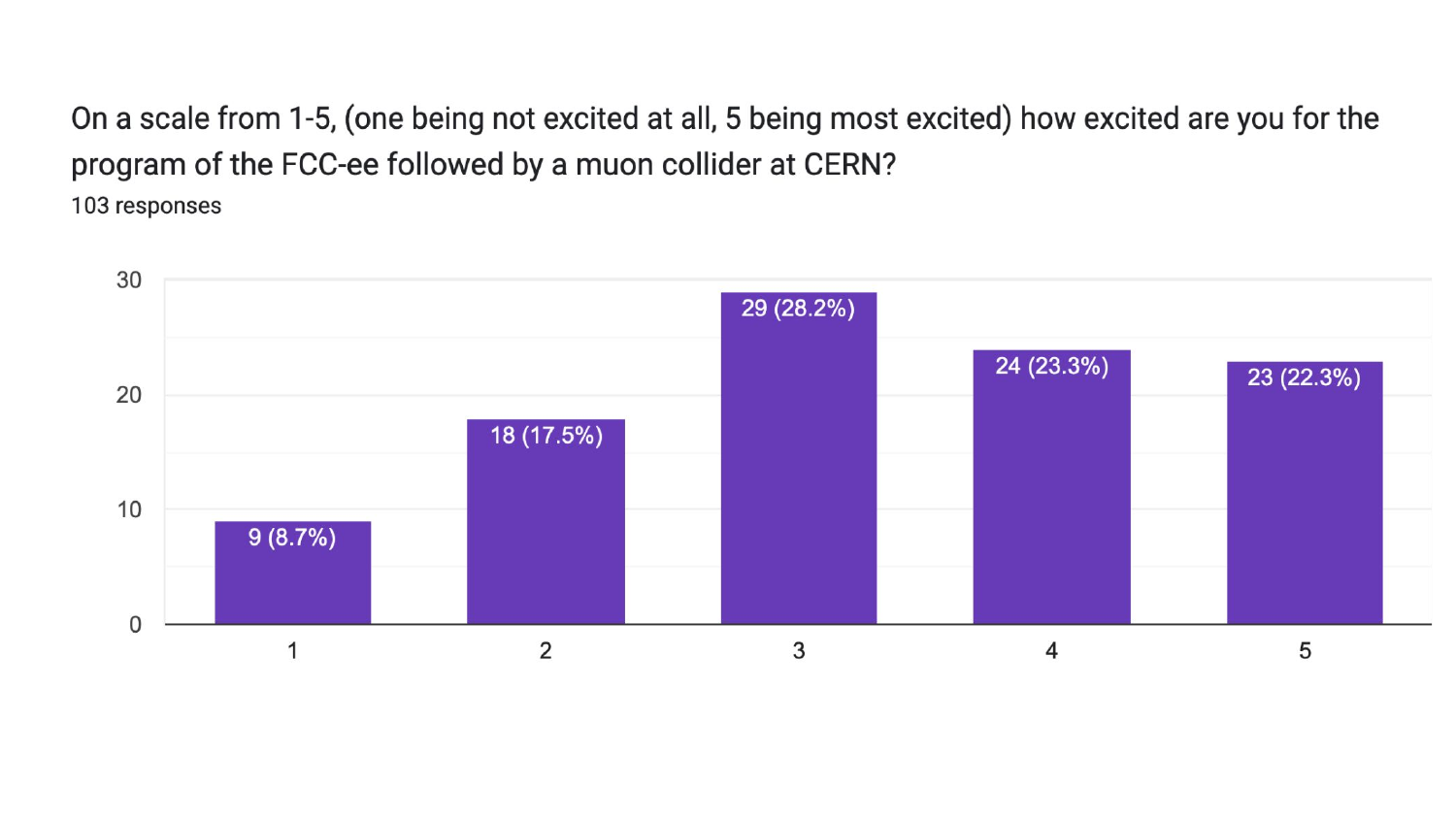}
    \includegraphics[width=0.5\linewidth]{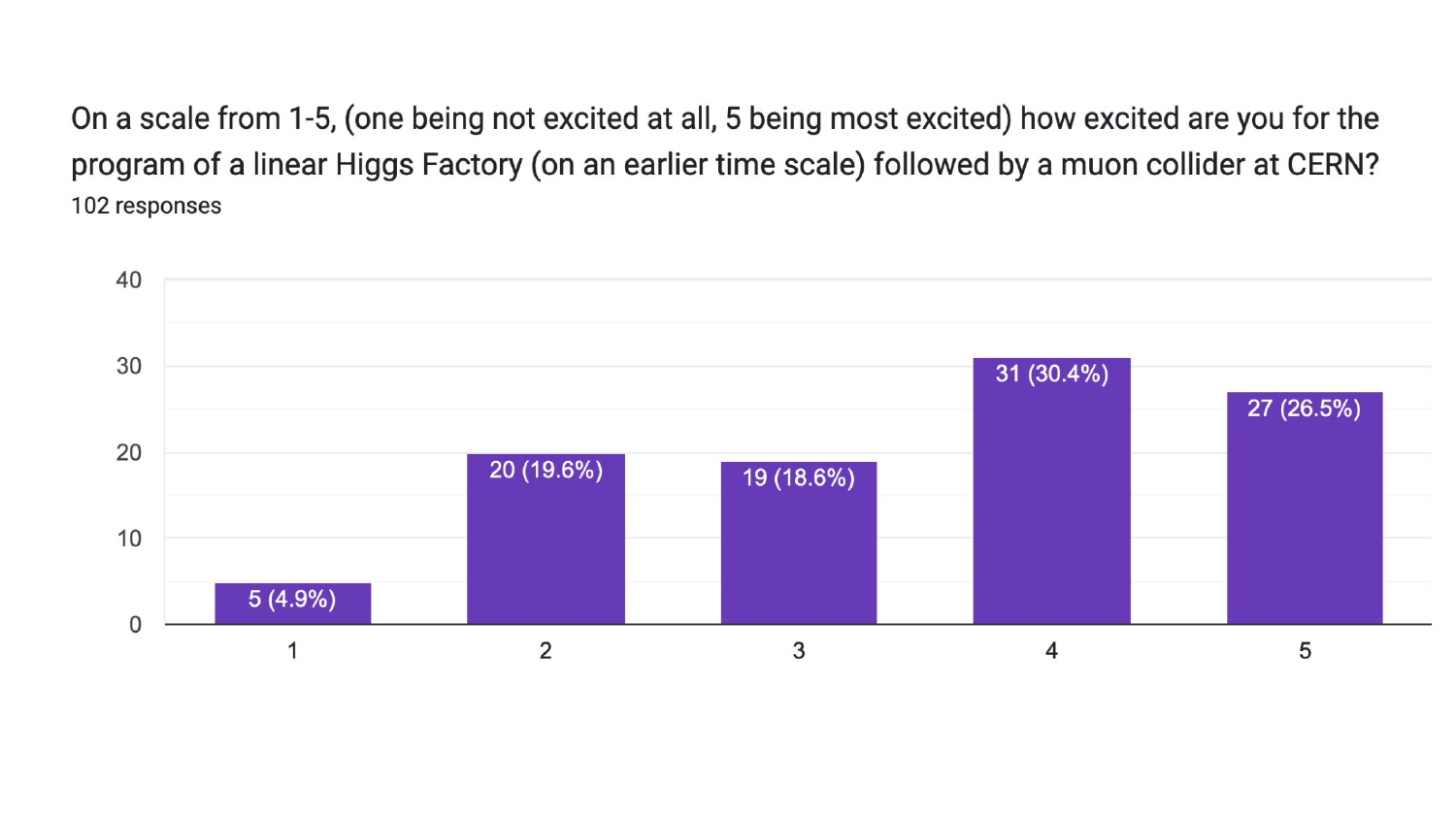} 
    \caption{Excitement for different project-staging combinations. Respondents were asked to rate their excitement from 1 (not excited at all) to 5 (very excited).}
    \label{fig:excite_combos}
\end{figure}

\begin{figure}[h]
    \centering
    \includegraphics[width=0.7\linewidth]{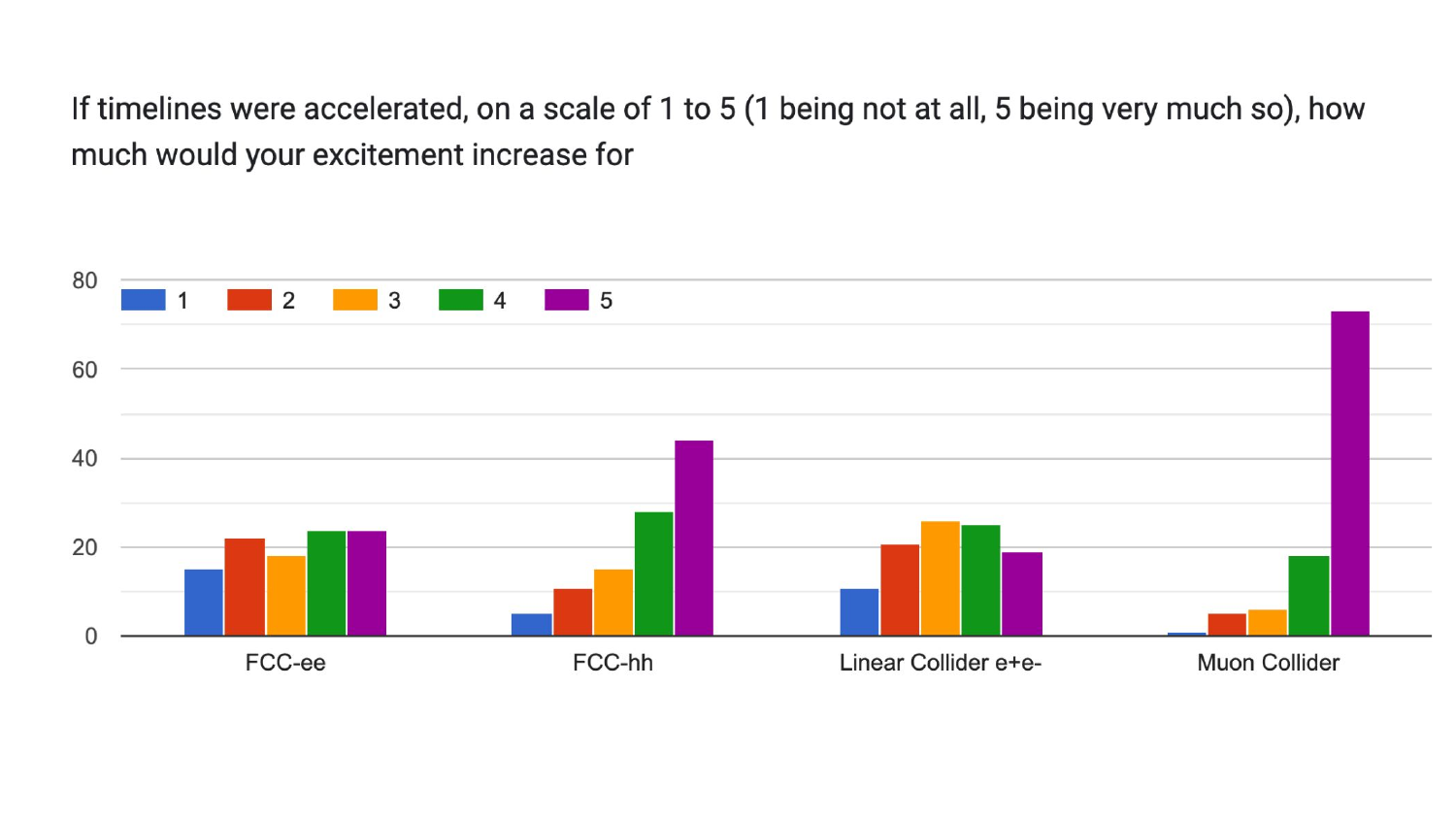}
    \includegraphics[width=0.7\linewidth]{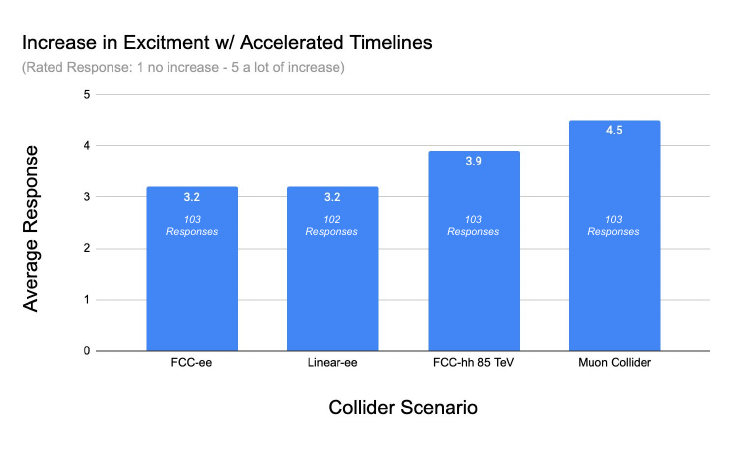}
    \caption{Participants were asked to indicate how much their enthusiasm would increase from 1 (not at all) to 5 (greatly) if project timelines were accelerated. Top: Detailed breakdown of responses received. Bottom: Average results per collider scenario.}
    \label{fig:excite_accel}
\end{figure}

\subsection{Detailed Preference Figures}
\label{sec:figs_addpref}
\begin{figure}[h]
    \centering
    \includegraphics[width=0.49\linewidth]{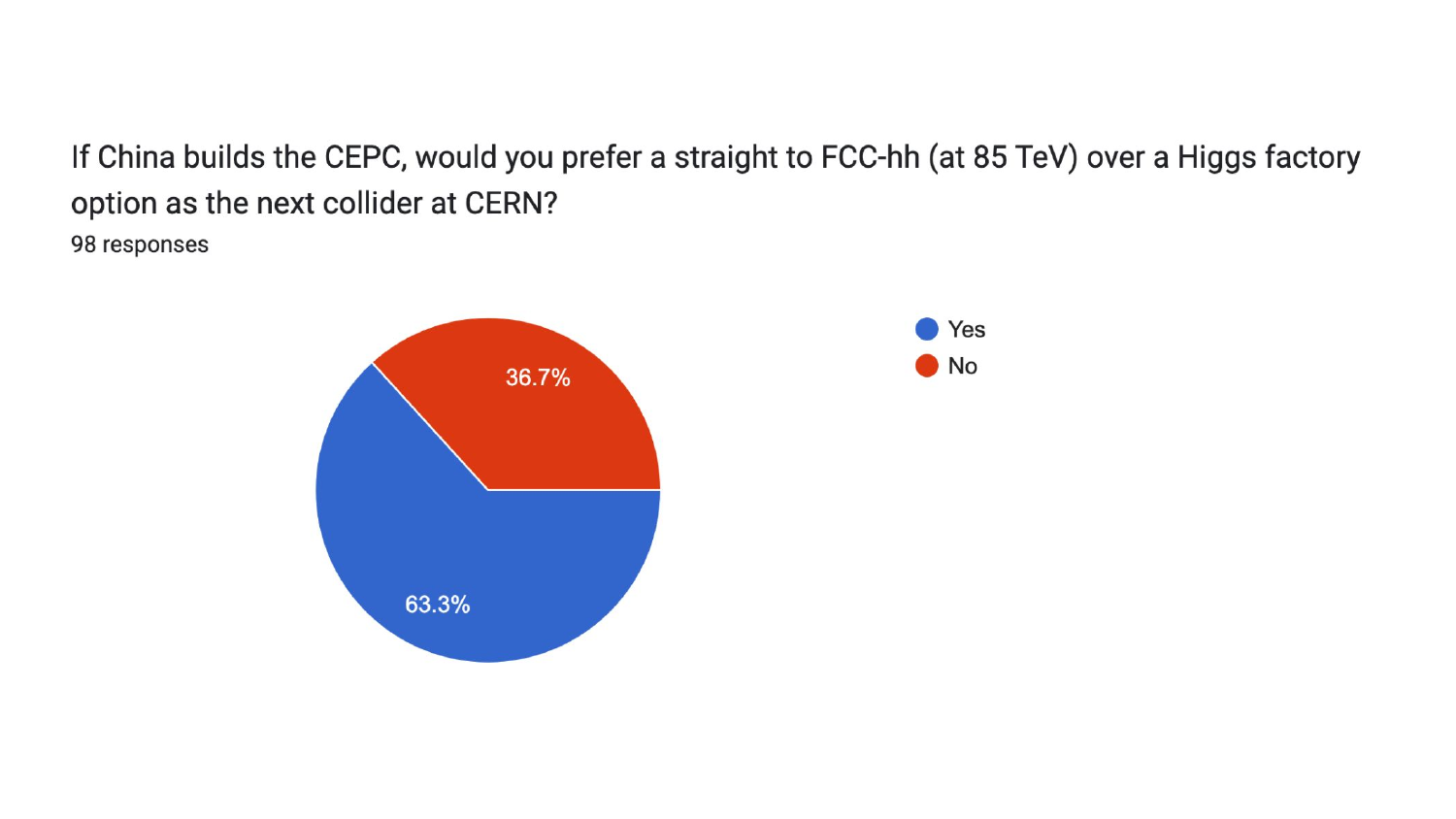}
    \includegraphics[width=0.49\linewidth]{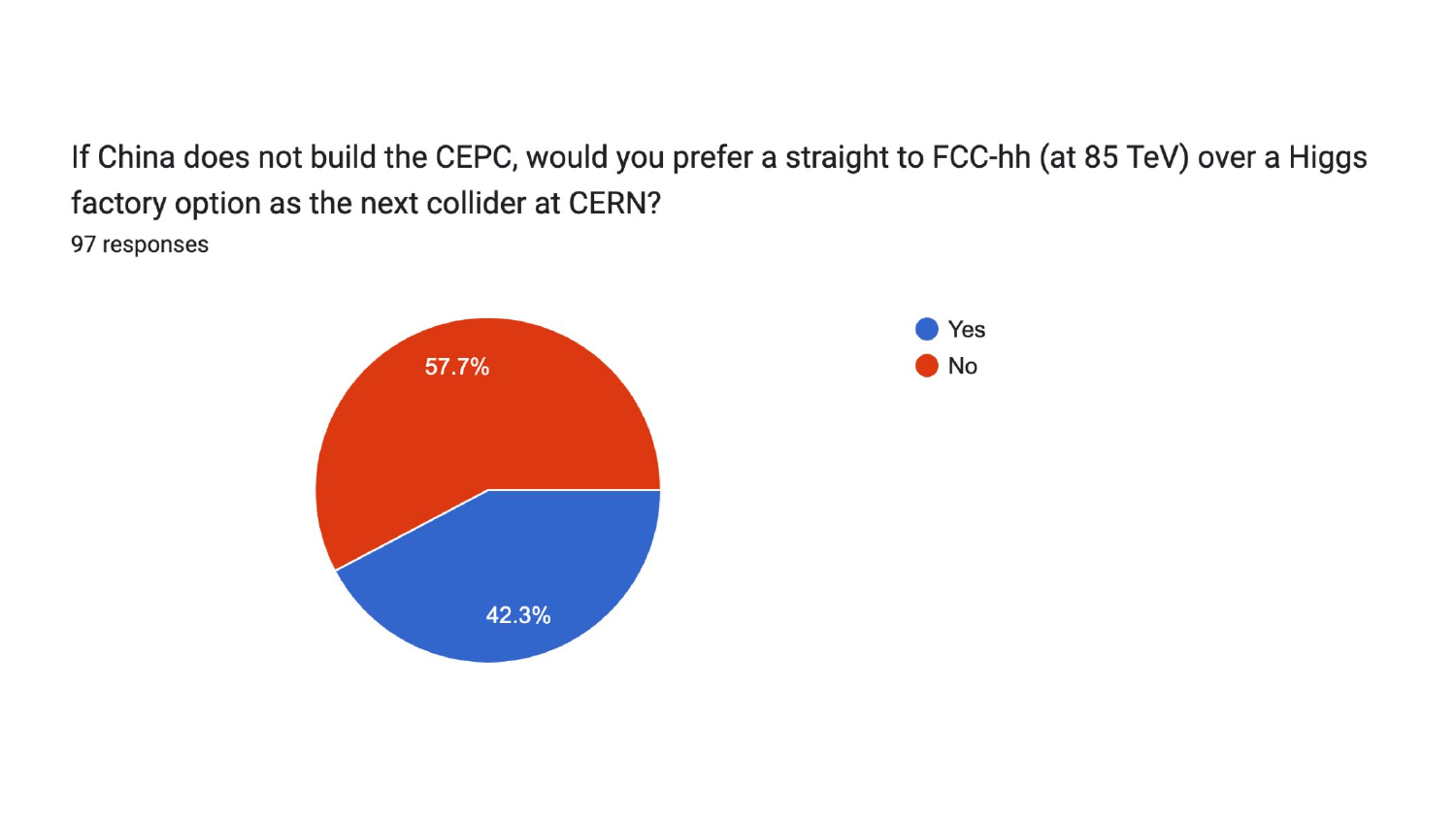}
    \caption{Summary of preferences for CERN's response with regards to the construction of the CEPC.}
    \label{fig:prio_cepc}
\end{figure}

\begin{figure}[h]
    \centering
    \includegraphics[width=0.6\linewidth]{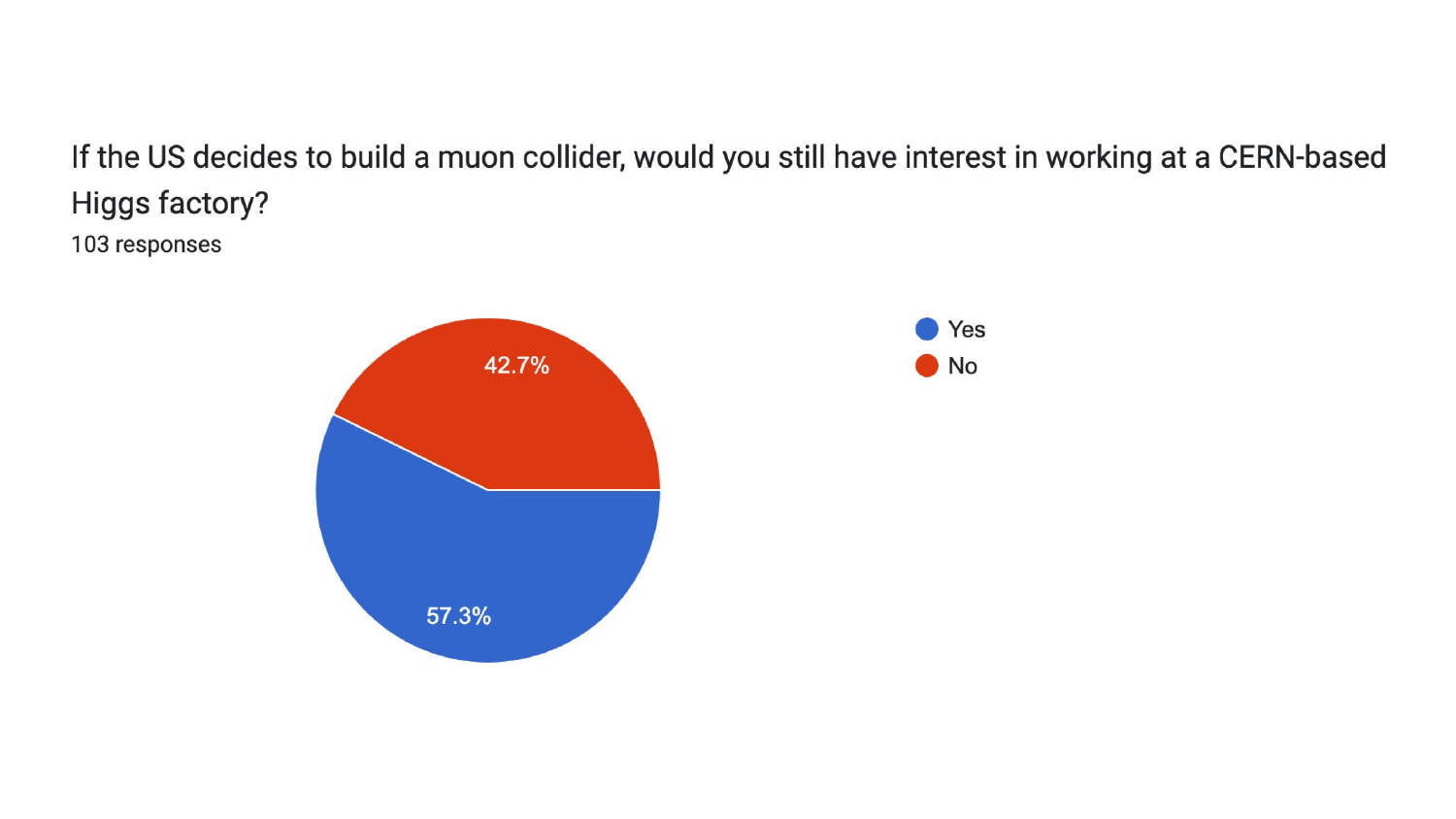}
    \includegraphics[width=0.6\linewidth]{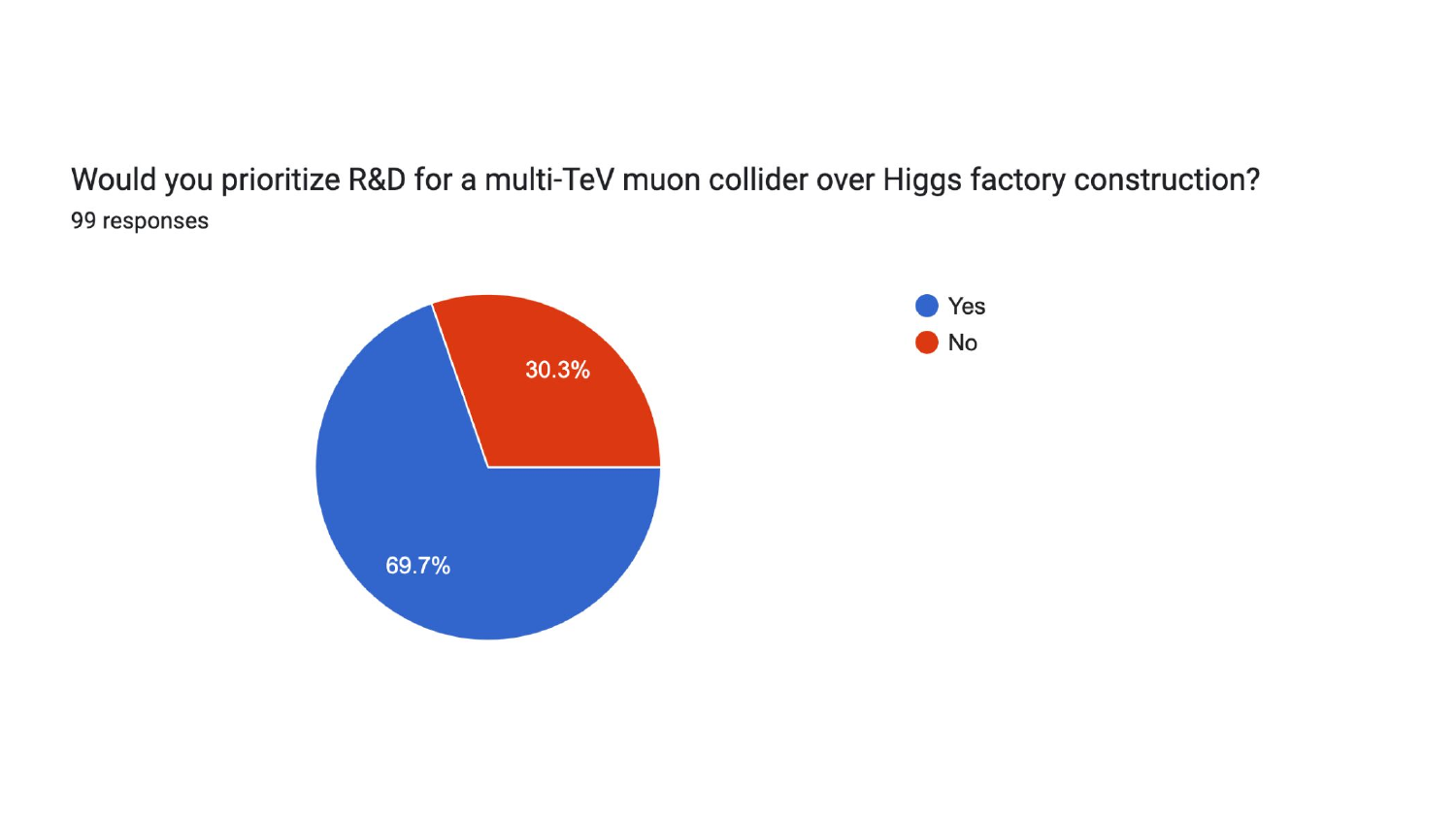}
    \includegraphics[width=0.6\linewidth]{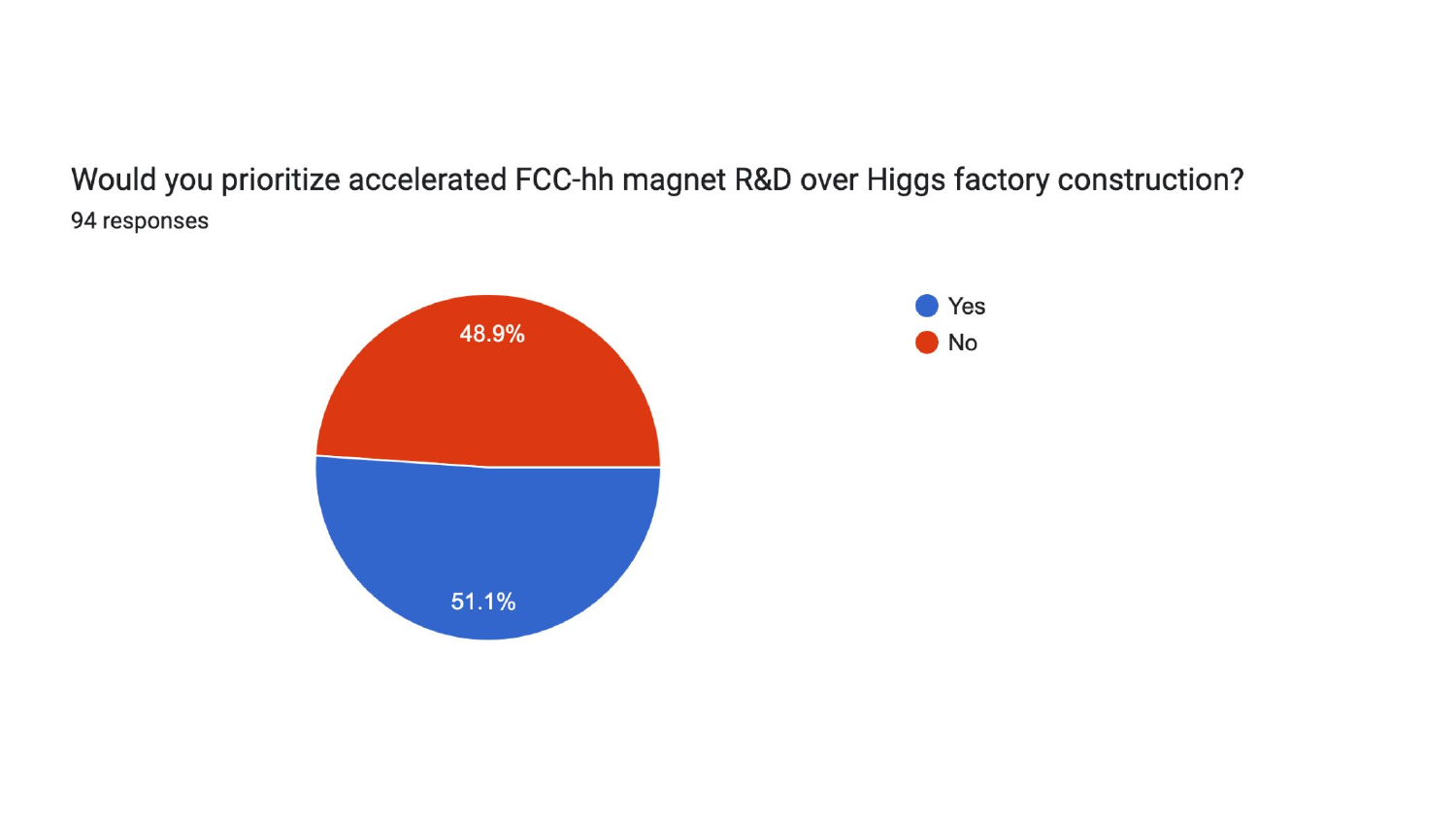}
    \caption{Summary of preferences regarding parton-center-of-mass colliders at 10 TeV.}
    \label{fig:prio_pcm10}
\end{figure}

\begin{figure}[h]
    \centering
    \includegraphics[width=0.7\linewidth]{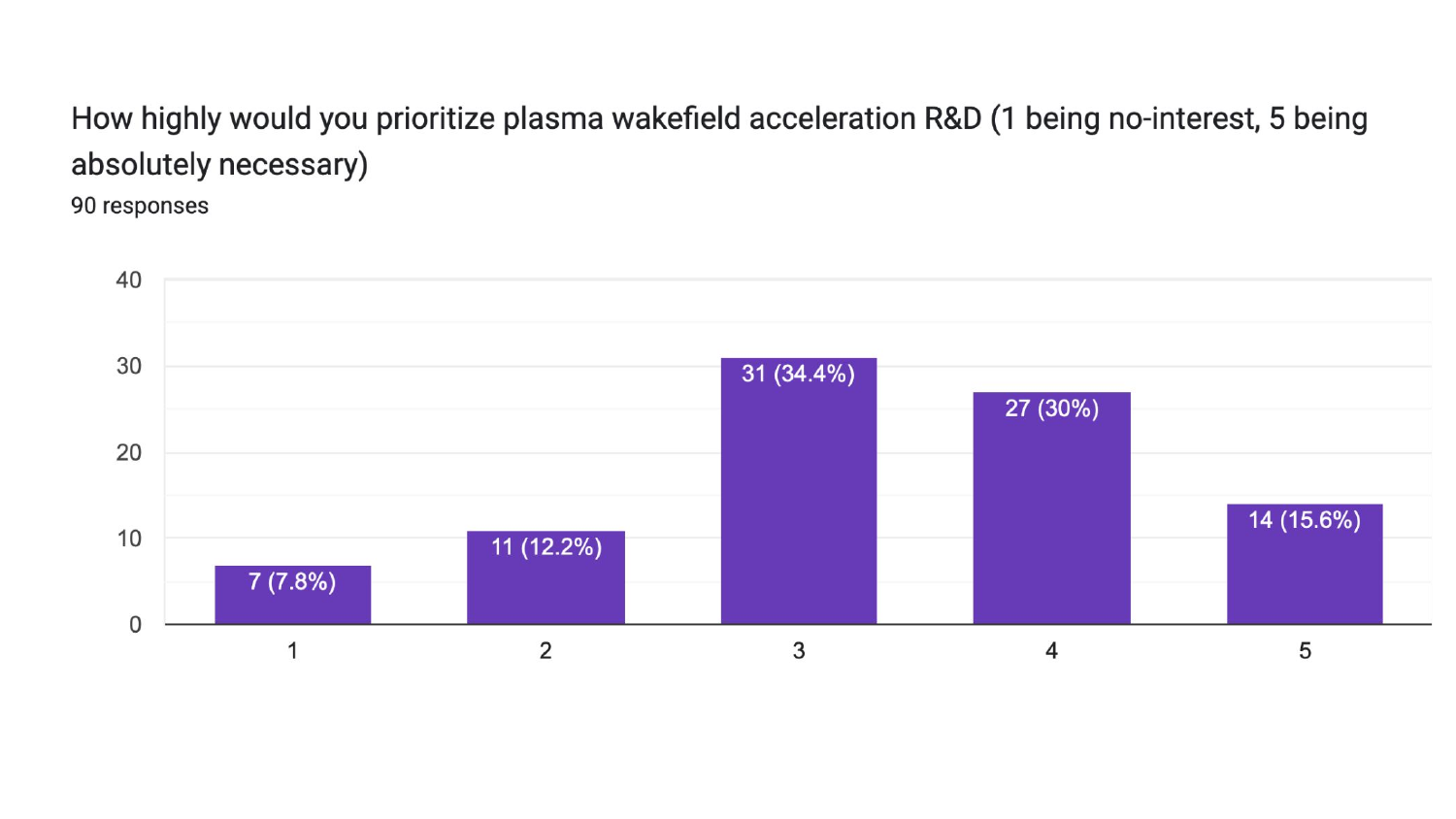}
    \caption{Detailed response to plasma-wakefield acceleration prioritization (1 not important at all, 5 very important.}
    \label{fig:prio_plasma}
\end{figure}
\clearpage
\subsection{Detailed Ranking Figures}
\label{sec:figs_addrank}

\begin{figure}[h]
    \centering
    \includegraphics[width=0.95\linewidth]{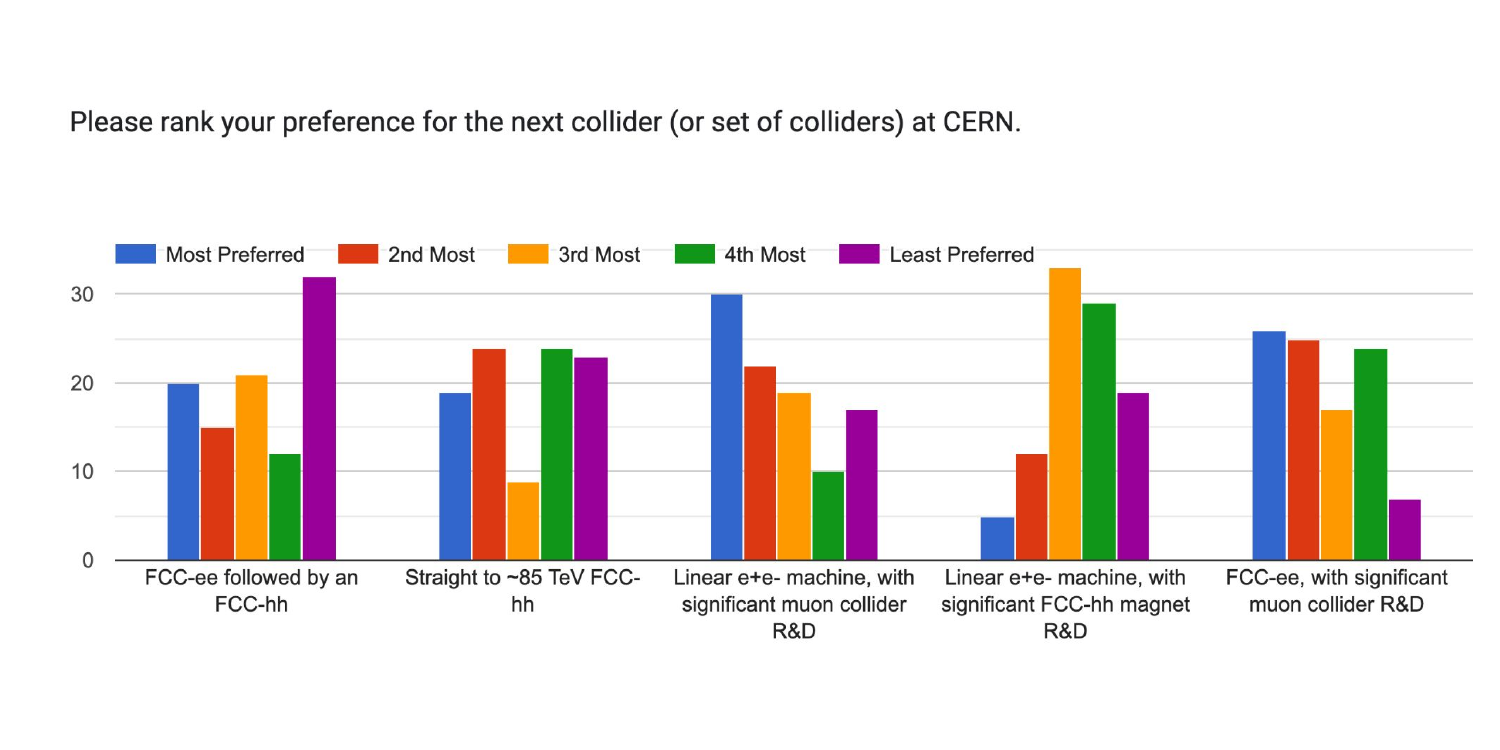}
    \caption{Rankings of different collider options.}
    \label{fig:rankings}
\end{figure}

\begin{figure}[h]
    \centering
    \includegraphics[width=0.7\linewidth]{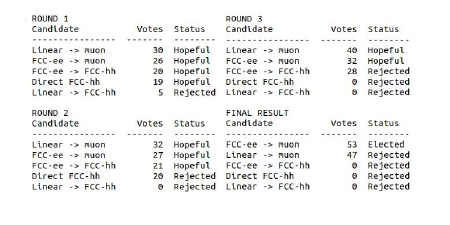}
    \caption{Results of a ranked choice voting algorithm applied to the rankings of the different collider options.}
    \label{fig:voting}
\end{figure}
\end{document}